\documentclass[,preprint,aps,prc,groupedaddress,showpacs]{revtex4}
\usepackage{graphicx}
\usepackage{hyperref}
\hypersetup{
    bookmarks=true,         % show bookmarks bar?
    unicode=false,          % non-Latin characters in Acrobat’s bookmarks
    pdftoolbar=true,        % show Acrobat’s toolbar?
    pdfmenubar=true,        % show Acrobat’s menu?
    pdffitwindow=false,     % window fit to page when opened
    pdfstartview={FitH},    % fits the width of the page to the window
    pdftitle={My title},    % title
    pdfauthor={Author},     % author
    pdfsubject={Subject},   % subject of the document
    pdfcreator={Creator},   % creator of the document
    pdfproducer={Producer}, % producer of the document
    pdfkeywords={keyword1} {key2} {key3}, % list of keywords
    pdfnewwindow=true,      % links in new window
    colorlinks=false,       % false: boxed links; true: colored links
    linkcolor=red,          % color of internal links (change box color with linkbordercolor)
    citecolor=green,        % color of links to bibliography
    filecolor=magenta,      % color of file links
    urlcolor=cyan           % color of external links
}
\def\eqref#1{Eq.~(\ref{eq:#1})}
\def\eqlab#1{\label{eq:#1}}
\def\figref#1{Fig.~(\ref{fig:#1})}
\def\figlab#1{\label{fig:#1}}
\def\tabref#1{Tab.~(\ref{tab:#1})}
\def\tablab#1{\label{tab:#1}}

\begin{document}

%\preprint{}

\title{Towards a model-independent constraint of the high-density dependence of the symmetry energy}


\author{M.D. Cozma}
\email[Corresponding author: ]{cozma@niham.nipne.ro}
%\thanks{}
\affiliation{IFIN-HH, Reactorului 30, 077125 M\v{a}gurele-Bucharest, Romania}

\author{Y. Leifels}
%\email[]{w.trautmann@gsi.de}
\affiliation{GSI Helmholtzzentrum f\"{u}r Schwerionenforschung GmbH, Planckstrasse 1, 64291 Darmstadt, Germany}

\author{W. Trautmann}
\affiliation{GSI Helmholtzzentrum f\"{u}r Schwerionenforschung GmbH, Planckstrasse 1, 64291 Darmstadt, Germany}

\author{Q. Li}
\affiliation{School of Science, Huzhou Teachers College, 313000 Huzhou, China}

\author{P. Russotto}
\affiliation{INFN-Sezione di Catania, 95123 Catania, Italy} 

\date{\today}

\begin{abstract}
Neutron-proton elliptic flow difference and ratio have been shown to be promising observables in the attempt
to constrain the density dependence of the symmetry energy above the saturation point from heavy-ion collision data.
Their dependence on model parameters like microscopic nucleon-nucleon cross-sections, compressibility of nuclear
matter, optical potential, and symmetry energy parametrization is thoroughly studied. 
By using a parametrization of the symmetry energy derived from the momentum dependent Gogny force in conjunction
with the T\"{u}bingen QMD model and comparing with the experimental FOPI/LAND data for $^{197}$Au+$^{197}$Au collisions at
400 MeV/nucleon, a moderately stiff ($L_{sym}$=122$\pm$57 MeV and $K_{sym}$=229$\pm$363 MeV) symmetry energy is extracted, a result
that agrees with that of a similar study that employed the UrQMD transport model and a power-law parametrization of the symmetry energy.
This contrasts with diverging results extracted from the FOPI $\pi^{-}/\pi^{+}$ ratio available in the literature.
\end{abstract}

%\keywords{}

\maketitle

\section{Introduction}
The isovector part of the equation of state (asy-EoS) of asymmetric nuclear matter, known as symmetry energy (SE),
represents one of the remaining open questions of nuclear physics. It comprises the lesser known part of the
nuclear matter equation of state (EoS) that can be approximately described by the expansion
\begin{eqnarray}
 E(\rho,\beta)&=&E_0(\rho,\beta=0)+E_{sym}(\rho)\,\beta^2\,,
\end{eqnarray}
where $\beta$=($\rho_n$-$\rho_p$)/($\rho_n$+$\rho_p$) with $\rho_n$, $\rho_p$ and $\rho$ denoting the neutron, proton
and total nucleon densities, respectively. The coefficient $E_{\rm sym}(\rho)$ of the asymmetry-dependent term is 
the symmetry energy. Knowledge of its precise density dependence is mandatory
for the proper understanding of the structure of rare isotopes, dynamics and spectra of heavy-ion collisions and
most importantly for certain astrophysical processes such as neutron star cooling and supernovae explosions~
\cite{Baran:2004ih,Li:2008gp}.
Intermediate energy nuclear reactions involving stable and radioactive beams have allowed by studying the thickness of neutron skins,
deformation, binding energies and isospin diffusion, the extraction of constaints on the density dependence of SE
at densities below saturation ($\rho_0$)~\cite{Li:1997px,Chen:2004si,Chen:2005ti,Tsang:2012se}. 
Existing theoretical models describing its density dependence generally agree with each other in this density regime, 
but their predictions start to diverge well before regions with densities $\rho\geq2\rho_0$ are reached~\cite{Li:2008gp}. 

Nuclear matter at suprasaturation densities is created in the laboratory in the processes of collisions of heavy nuclei.
Several observables that can be measured in such reactions have been determined to bear information on the behavior
of the SE above $\rho_0$: the ratio of high transverse momentum neutron/proton yields~\cite{Yong:2007tx}, 
light cluster emission~\cite{Chen:2003qj}, $\pi^-/\pi^{+}$ multiplicity ratio in central collisions~
\cite{Xiao:2009zza,Feng:2009am,Xie:2013np},
 double neutron to proton ratios of nucleon emission from isospin-asymmetric but
mass-symmetric reactions~\cite{Li:2006wc} and others. 

The FOPI experimental data for the $\pi^-/\pi^{+}$ 
ratio~\cite{Reisdorf:2006ie} have been used to set constraints on the suprasaturation density behavior of SE 
by various authors with contradicting results: Xiao $et\,al.$~\cite{Xiao:2009zza} made use of the IBUU transport model
supplemented by the isovector momentum dependent Gogny inspired parametrization of  symmetry potential~\cite{Das:2002fr} to
point toward a soft asy-EoS, the study of Feng and Jin~\cite{Feng:2009am}, which employed 
the isospin-dependent quantum molecular dynamics (IQMD) model and a power-law parametrization of the symmetry energy,
\begin{eqnarray}
 S(\rho)&=&S_0\,(\rho/\rho_0)^{\gamma},
\eqlab{sepowerlaw}
\end{eqnarray}
favors a stiff SE. Most recently Xie $et\,al.$~\cite{Xie:2013np} addressed the same issue within the
Boltzmann-Langevin approach and a power-law
parametrization of asy-EoS presenting support for a super-soft scenario for the symmetry energy.

Constraints on the high density dependence of $S(\rho)$ extracted from elliptic flow  ratios of neutrons and
protons (npEFR) $v_2^{n/p}=v_2^n/v_2^p$ and of neutrons and hydrogen $v_2^{n/H}$ have been presented by 
Russotto $et\,al.$~\cite{Russotto:2011hq}. The experimental data taken
by the FOPI-LAND Collaboration for $^{197}$Au+$^{197}$Au collisions at 400 MeV/nucleon incident energy~
\cite{Leifels:1993ir,Lambrecht:1994cp} have been reanalyzed allowing for a reduction of systematical and statistical uncertainties. 
To model heavy-ion collisions a version of the UrQMD model~\cite{Li:2005gfa,Li:2005zza} and the power-law parametrization
of SE mentioned above have been employed. A comparison of the theoretical and experimental elliptic flow ratios of 
neutrons vs. protons ($v_2^n/v_2^p$) and neutrons vs. hydrogen ($v_2^n/v_2^H$) has led to a constraint compatible with a
linear density dependence for the potential part $S(\rho)$: $\gamma$=0.9$\pm$0.4~\cite{Russotto:2011hq}.

In an independent study~\cite{Cozma:2011nr} the neutron-proton elliptic flow difference (npEFD) $v_2^{n-p}=v_2^n-v_2^p$ has been proposed
as a viable observable for constraining the suprasaturation density dependence of SE. Its dependence on
model parameters like in-medium microscopic nucleon-nucleon cross-sections, compressibility of symmetric nuclear matter
and width of the gaussian wave packet of nucleons is a rather small fraction of the sensitivity to the changes between a stiff and a soft
asy-EoS for kinematical acceptances close to those of the FOPI experiment. A comparison with published FOPI-LAND impact
parameter dependent data~\cite{Leifels:1993ir,Lambrecht:1994cp} for $v_2^{n-p}$ was found problematic due to a highly non-monotonous 
dependence of the experimental data on that variable. Still, the experimental $v_2^{n-H}$, viewed as an upper bound of $v_2^{n-p}$,
allowed the exclusion of the super-soft asy-EoS from the list of possible scenarios.

The present Article aims at an update of the results of References~\cite{Cozma:2011nr,Russotto:2011hq} by extending
the analysis of the former to both neutron-proton elliptic flow differences $v_2^{n-p}$ and ratios $v_2^{n/p}$ and by addressing
the model dependence due to the momentum dependent part of the EoS and of the momentum dependence of the symmetry potential. 
A recent overview addressing the relevance of elliptic flow in the study of the SE at supra-saturation density can be found 
in Ref.~\cite{Trautmann:2012nk}.

\section{The framework}
\subsection{The model}
In the present study, heavy-ion collisions have been simulated by using the QMD transport model developed
in T\"{u}bingen~\cite{Khoa:1992zz,UmaMaheswari:1997ig} and expanded to accommodate density-dependent
nucleon-nucleon cross-sections and an isospin dependent EoS. The same model has been previously used to study 
dilepton emission in heavy-ion collisions~\cite{Shekhter:2003xd,Cozma:2006vp,Santini:2008pk}, stiffness 
of the equation of state of symmetric nuclear matter~\cite{Fuchs:2000kp} and various in-medium effects relevant for the dynamics of heavy-ion 
collisions~\cite{Fuchs:1997we,UmaMaheswari:1997ig}. Most of the results of the following Section
have been obtained by making use of the Gogny inspired momentum dependent parametrization of the isovector
part of the equation of state~\cite{Das:2002fr}. It contains a  parameter denoted $x$ which has been
introduced to allow adjustments in the stiffness of asy-EoS, negative and positive values corresponding to a stiff
and a soft density dependence of the symmetry energy, respectively (see Sect. IIIA). To assess the importance of the momentum dependent  
part of asy-EoS, the momentum-independent power-law parametrization (\eqref{sepowerlaw}) is used where indicated. 
Further details of the model, relevant for the current study, can be found in~\cite{Cozma:2011nr}. 
\subsection{Experimental data}

\begin{table*}[t]
 \centering
\begin{tabular}{|c|c|c|c|c|c|}
\hline\hline
Data set&Centrality&b (fm)&$ v_2^n$ & $v_2^p$ &$v_2^H$\\
\hline\hline
&E2&7.2&-0.0939$\pm$0.0059& -0.0966$\pm$0.0052&-0.1045$\pm$0.0040\\
B&E3&4.7&-0.0711$\pm$0.0057&-0.0705$\pm$0.0054&-0.0758$\pm$0.0040\\
$0.25\leq y/y_P\leq 0.75$&E4&3.4&-0.0615$\pm$0.0066&-0.0324$\pm$0.0066&-0.0501$\pm$0.0047\\
$0.3\leq p_T\leq 1.0$&E5&1.9&-0.0245$\pm$0.0068&-0.0201$\pm$0.0072&-0.0222$\pm$0.0049\\
&E2-E5&$<$7.5&-0.0655$\pm$0.0031&-0.0627$\pm$0.0030&-0.0681$\pm$0.0022\\
\hline
&E2&7.2&-0.1065$\pm$0.0111&-0.1008$\pm$0.0101&-0.1249$\pm$0.0079\\
C&E3&4.7&-0.0681$\pm$0.0108&-0.0583$\pm$0.0110&-0.0841$\pm$0.0080\\
$0.45\leq y/y_P\leq 0.55$&E4&3.4&-0.0552$\pm$0.0125&-0.0356$\pm$0.0129&-0.0593$\pm$0.0090\\
$0.3\leq p_T\leq 1.0$&E5&1.9&-0.0259$\pm$0.0126&0.0007$\pm$0.0148&-0.0178$\pm$0.0095\\
&E2-E5&$<$7.5&-0.0668$\pm$0.0058&-0.0586$\pm$0.0059&-0.0771$\pm$0.0043\\
\hline\hline
\end{tabular}
\caption{Experimental FOPI-LAND values for elliptic flow of neutrons, protons and hydrogen
for two choices of kinematical conditions referred to in the text as data sets B and C. The applied kinematical cuts are shown
in the first column, the values of the transverse momentum $p_T$ are in units of GeV/c. Changes in the elliptic
flow values are minute (fourth digit) when the transverse momentum cut is relaxed to $0.3 \leq p_{T} \leq 1.3$.}
%The corresponding values for data set A can be found in Ref.~\cite{Lambrecht:1994cp}.}
\tablab{expdata}
\end{table*}

In the original release of the FOPI-LAND data~\cite{Leifels:1993ir,Lambrecht:1994cp} the extraction of proton spectra required,
due to insufficient calorimeter resolution, narrow constraints to be applied in order to minimize the contamination with 
deuterium and tritium events.
% conservative cuts to be applied in order to eliminate deuterium and
%tritium events. 
As a result proton elliptic flow values show a non-monotonous dependence on the impact parameter, in
contrast to neutrons, making a comparison with predicted values troublesome. The data have been recently reanalyzed
~\cite{Russotto:2011hq} in order to determine the optimum conditions for the new ASY-EOS experiment~\cite{Russotto:2012jb}
and to extract constraints on the density dependence of the symmetry energy. 
This effort has also resulted in a smoother impact-parameter dependence of the elliptic flow results for protons. 
Several data sets, corresponding to different ranges in
rapidity $0.25<y/y_P<0.75$ (B) and $0.45<y/y_P<0.55$ (C) and transverse momentum
\footnote{Units of momenta, GeV/c, are not displayed explicitly.} 
($0.3<p_T<1.0$ and $0.3<p_T<1.3$) are available. The experimental values for the elliptic flow of neutrons and protons,
as well as hydrogen, used to obtain the results in this work are displayed in~\tabref{expdata}.
In the original FOPI-LAND data~\cite{Leifels:1993ir,Lambrecht:1994cp}, only the rapidity window $0.40 < y/y_p < 0.60$ (A)
is covered for which the kinematical acceptance of the FOPI-LAND detector produces constraints for the transverse 
momentum as well: $0.27 < p_T < 1.06$. The values of the elliptic flow of neutrons and protons for this data set have
been derived using the experimental values for the squeeze-out factor $R_N$ presented in Ref.~\cite{Lambrecht:1994cp}.

Constraints on the stiffness of asy-EoS extracted by a comparison of the data sets corresponding to different rapidity
windows, integrated over impact parameter, with model predictions of EFD $v_2^{n-p}$ agree with each other. 
One obtains for the $x$ parameter the following values:  $x=-2.5\pm1.5$, $x=-1.5\pm0.75$, and $x=-2.0\pm0.75$
for the data sets A, B and C, respectively. Differences between the choices $p_T<1.0$ and $p_T<1.3$ are negligible. 
The theoretical estimates were obtained using the set of model
parameters employed in Section III to generate the central solution, the uncertainty in the values of the $x$ parameter
originating solely from the error bars of experimental elliptic flow values.

\section{Model dependence}

\subsection{Parametrizations of the potential}
The nucleon optical potential is an important ingredient of transport models, the sensitivity of heavy-ion
observables in general~\cite{Aichelin:1987ti,Peilert:1989kr,Greco:1999zz} and collective flows~\cite{Zhang:1994hpa}
in particular to its momentum dependence being well documented. In a recent study~\cite{Zhang:2012fc}
the effects of the momentum dependence of the symmetry potential on transverse and elliptic flows have been investigated
with the conclusion that the neutron-proton elliptic flow difference exhibits a small sensitivity
to the momentum dependent part of the isovector nucleon potential within the constraint of an asy-soft EoS. This
is an important finding since the momentum dependence of the isovector part of the nucleon potential is still
an open question. Parametrizations of it with various momentum dependences, or none at all, are commonly employed.

On the theoretical side, the optical potential has been extracted from first principles~\cite{Baldo:1989zz}, and 
similar approaches have later been extended to also extract the symmetry potential~\cite{Zuo:2001bd,Zuo:2005hw,Xu:2012fu}.
Alternatively, it has been possible to extract the momentum dependence of the bare nucleon interaction within an effective
model~\cite{Hartnack:1994zz} starting from the optical potential of Refs.~\cite{Arnold:1982rf,Hama:1990vr} obtained within
a relativistic Dirac-equation description of experimental data of proton scattering on Ca and heavier nuclei. The results
of the two approaches are somewhat different, the Brueckner-Hartree-Fock approach and its
relativistic counterpart favor a potential that is attractive at all values of the momentum, while the relativistic Dirac approach
delivers a potential that becomes repulsive above a certain momentum threshold depending on which experimental data sets
are considered. Additionally, the Brueckner-Hartree-Fock approach predicts an optical potential that is almost momentum
independent at moderate values of the momentum.

To account for this model dependence we have simulated heavy-ion collisions by considering three different parametrizations
of the optical potential. The first one stems from the isoscalar part of the Gogny interaction~\cite{Das:2002fr} while
the last two mimic the parametrizations presented in Ref.~\cite{Hartnack:1994zz}
\begin{eqnarray}
 V_{opt}^{(MDI)}(\vec{p_i},\vec{p_j})&=&(C_l+C_u)\,\frac{1}{1+(\vec{p_i}-\vec{p_j})^2/\Lambda^2}\,\frac{\rho_{ij}}{\rho_0} \nonumber\\
  V_{opt}^{(HA)}(\vec{p_i},\vec{p_j})&=&\{V_0+v\,\mathrm{ln}^2[\,a\,(\vec{p_i}-\vec{p_j})^2+1]\}\,\frac{\rho_{ij}}{\rho_0}.
\eqlab{optpotparam}
\end{eqnarray}
The parameters present in $V_{opt}^{(MDI)}(\vec{p_i},\vec{p_j})$ can be found in Ref.~\cite{Das:2002fr}, while
for $V_{opt}^{(HA)}(\vec{p_i},\vec{p_j})$ they read: $V_0$=-0.054 GeV, $v$=0.00158 GeV and $a$=500 GeV$^{-2}$,
$V_0$=-0.0753 GeV, $v$=0.002526 GeV and $a$=500 GeV$^{-2}$ for the old and new parametrization in 
Ref.~\cite{Hartnack:1994zz}, respectively; $\rho_{ij}$ is the contribution to the density at the location of nucleon $j$ due to nucleon $i$,
recovering in the infinite nuclear matter limit the parametrization of Ref.~\cite{Das:2002fr} and the right EoS. The
$V_0$ parameter is absorbed in the linearly density dependent term of the single nucleon potential $V=\alpha\,\frac{\rho}{\rho_0}
+\beta\,(\frac{\rho}{\rho_0})^\gamma+V_{opt}$. For completeness, the values of the remaining parameters, producing
a soft (K=210 MeV) isoscalar EoS, read:
$\alpha$=-0.3901 GeV, $\beta$=0.3203 GeV, $\gamma$=1.14 and $\alpha$=-0.2017 GeV, $\beta$=0.1861 GeV, $\gamma$=1.2104
for the two $HA$ parametrizations.

The momentum dependence of the symmetry potential is currently an unsettled issue and consequently various parametrizations
have been employed in the literature. To estimate the impact of this unknown on elliptic flow observables, we have selected
two of the most widely employed parametrizations for the current study: the Gogny interaction inspired one (Refs.~\cite{Chen:2004si, Das:2002fr}),
producing a momentum dependent symmetry potential,
\begin{eqnarray}
&&V(\rho,\beta)=\frac{A_1}{2\rho_0}\rho^2+\frac{A_2(x)}{2\rho_0}\rho^2\beta^2
+\frac{B}{\sigma+1}\frac{\rho^{\sigma+1}}{\rho_0^\sigma}(1-x\beta^2) \nonumber\\
&&\phantom{aaaa} +\frac{1}{\rho_0}\sum_{\tau,\tau'}\,C_{\tau\tau'}\int\int d^3 p d^3 p'\,\frac{f_{\tau}(\vec{r},\vec{p})\,f_{\tau'}(\vec{r},\vec{p'})}
{1+(\vec{p}-\vec{p}')^2/\Lambda^2}
\eqlab{sympot}
\end{eqnarray}
and the power-law parametrization, that leads to a momentum independent potential
\begin{eqnarray}
S(\rho)=\left\{
\begin{array}{l}
 S_0\,(\rho/\rho_0)^\gamma \hspace{0.25cm} \text{-  linear, stiff} \\ 
a+(S_0-a)(\rho/\rho_0)^\gamma  \hspace{0.25cm} \text{-  soft, supersoft}.
\end{array}
\right.
\eqlab{sympotpowerlaw}
\end{eqnarray}
with $S_0$=18.5 MeV. 

To reproduce different density dependencies of the symmetry energy predicted by various ab-initio theoretical 
calculations the original Gogny interaction has been generalized in Ref.~\cite{Das:2002fr} by introducing a real parameter $x$
that can be adjusted to generate an asy-EoS with the desired saturation density magnitude and high density behavior. Values
of the parameters present in~\eqref{sympot} for the choices $x=1$ (soft) and $x=0$ (close to linear) can be found in
Ref.~\cite{Das:2002fr} for the case of a soft iso-scalar EoS ($K$=210 MeV). Reproduction of the saturation value
of the symmetry energy requires that the parameter $A_2$ bears a linear dependence on $x$. Consequently, for a given
density value, the symmetry energy's dependence on $x$ is linear. This implies that the coefficients of the Taylor expansion
of the SE around saturation density, in particular $L_{sym}$ and $K_{sym}$ defined in~\eqref{lsymksym}, bear a linear dependence
on $x$. Stiff and soft SE density dependencies can be simulated by choosing negative and positive values for $x$, respectively.

In the case of the power-law parametrization of SE,~\eqref{sympotpowerlaw}, density dependencies close to those provided
by the Gogny interaction for $x$=-2.0,-1.0, 0.0 are obtained for values of the parameter $\gamma$=2.0, 1.5, 0.5, respectively.
To mimic the soft and super-soft symmetry energy provided by the Gogny interaction with
$x$=1 and respectively $x$=2 modified power-law parametrizations, as presented in ~\eqref{sympotpowerlaw},
are employed above the saturation point. The sets of parameters $a=23.0$ MeV, $\gamma$=1.0 and 
$a=31.0$ MeV, $\gamma$=2.0 reproduce a soft and super-soft density dependence respectively. Below saturation, the standard power-law 
parametrization with $\gamma$=1.0 is employed in these cases. The soft power-law parametrizations,
while producing discontinuous force terms at saturation density, have the advantage of having an identical functional density
dependence of the force term as its stiff counterpart and generating a SE stiffness below saturation point compatible with
the experimental result for that density region.

\subsection{Model dependence of observables}

Elliptic flows of protons and neutrons cannot be used separately to constrain the isovector part of the equation of state
above the saturation point due to their sizable dependence on particular values of transport model parameters, 
that are either inaccurately determined or do not represent measurable quantities, 
like in-medium nucleon-nucleon ($NN$) cross-sections, compressibility modulus of nuclear matter, 
width of the nucleon wave function and strength of the optical potential. A precise
knowledge of the dependence on these parameters would allow the elimination of the most uncertain ones leaving one
with a set of observables that bear no or almost no model dependence. In practice, one is forced to make assumptions that
are either verified or disproved by employing a definite transport model. Neutron proton elliptic flow ratios 
require a scenario in which elliptic flows scale linearly with model parameters, while for the differences the variation of elliptic flows
of neutrons and protons with respect to model parameters should be equal in order to be able to extract
a model independent constraint on the density dependence of SE. 

%\begin{turnpage}
\begin{figure}[t]
\includegraphics[width=20pc]{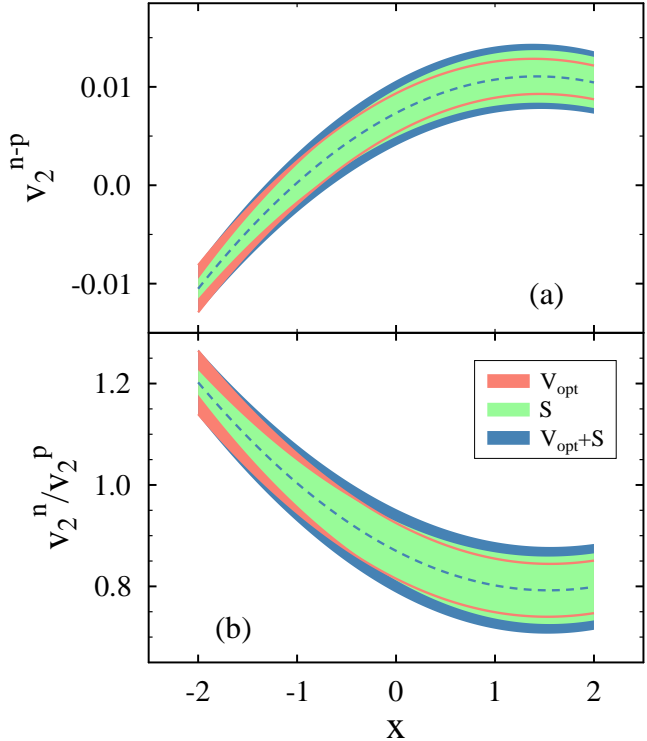}
\caption{\figlab{npefdoptpotsens}(Color online) Variations in the values of the impact parameter integrated ( b$\leq$7.5 fm ) npEFD (a) and
npEFR (b) due to different choices for the optical potential ($V_{opt}$), parametrization of symmetry-energy
($S$) as well as the combined, quadratically added, uncertainty.}
\end{figure}

%At this point two slightly different approaches are possible. The first one relies on one of the two observables, either npEFD or npEFR, being insensitive
%to model parameter variations and extract constraints on the stiffness of the asy-EoS employing that particular observable. In practice
%the total model dependencies of the two are comparable in magnitude and using the fact that in this approach npEFD and npEFR are independent
%observables the final results on the asy-EoS stiffness lies in the overlapping region of the constraints extracted using each observable independently.
%This approach does not impose at the beginning that experimental elliptic flow values are reproduced but the last step enforces that. 
%In the second approach one requires from the very beginning that the experimental elliptic flow values are reproduced for
%a particular scenario for the asy-EoS and in consequence npEFD and npEDR cannot be treated as independent observables anymore.
%Differences in the constraints extracted by using each of them independently are related to how close the experimental elliptic flow values are reproduced
%for the favored asy-EoS scenario. We will employ this approach in the following.
To extract constraints on the stiffness of the SE the following procedure is employed. We require that the experimental
elliptic flow values are reproduced as closely as possible for a particular asy-EoS scenario, while keeping model parameters
within limits commonly found in the literature. As a consequence, npEFD and npEDR cannot be treated as independent observables anymore.
Differences in the constraints extracted by using each of them independently are related to how close the experimental elliptic 
flow values are reproduced for the favored asy-EoS scenario.

In the following the sensitivity of npEFD and npEFR with respect to model parameters will be presented.
For the central estimates, a reproduction of the experimental elliptic flow data for a value of the asy-EoS stiffness parameter $x$=-1 
within 10-15$\%$ was possible with the following set of parameter values: stiffness of the isoscalar EoS set to $K$=210 MeV, 
width of the nucleon wave function $L$=4.33 fm$^2$, the new version of the $V_{opt}^{(HA)}$ optical potential parametrization 
and Cugnon nucleon-nucleon cross-sections. We would
like to note that very few of the possible combinations of model parameters produce values for the elliptic flow compatible with
the experimental values, many combinations underestimate its strength severely, sometimes up to a factor of 2. Once the stiffness
of the scalar part of the EoS is set to a soft one ($K$=210 MeV) the choices to enhance elliptic flow towards realistic values are
to decrease the nucleon wave function width and/or choose an optical potential that is as repulsive as possible in its higher energy region. 
The Cugnon nucleon-nucleon cross-section parametrization has been used in the collision term. 
It should be noted that Cugnon neutron-proton cross-sections are lower than the experimental vacuum ones below an incident kinetic energy
of $T$=100 MeV, but significantly higher than the in-medium theoretical predictions at saturation density.  They can, therefore, be thought of as
effectively simulating some in-medium effects.

Results for the sensitivity of npEFD and npEFR to both the momentum dependent part of the isospin symmetric EoS and 
various parametrizations of the symmetry energy are displayed in \figref{npefdoptpotsens} as a function of the stiffness of the asy-EoS. 
Collisions of Au+Au at 400 MeV/nucleon have been simulated and the kinematical cuts labeled 'B' above have been applied. 
The widths of the bands represent the variations of npEFD or npEFR  when switching between parametrizations of the optical
potential while keeping the SE parametrization fixed and vice-versa. The results presented correspond to averages over different
choices of the quantity that was kept fixed. Conclusions as $e.g.$ the sensitivity/insensitivity of the studied observables
to the momentum dependence of symmetry potential for an asy-soft scenario (as was presented in Ref.~\cite{Zhang:2012fc}) can thus not
be drawn from this figure. Each of the possible combinations of optical potential parametrization and symmetry energy
parametrization (6 in total) usually yields a different outcome in this respect. The result of~\figref{npefdoptpotsens} should therefore
be considered as an estimate. Nevertheless, it can be concluded that the uncertainties in the optical potential and the
momentum dependence or independence of the symmetry potential have an important impact on elliptic flow observables like npEFD and npEFR.
For precisely constraining the symmetry energy at high density from elliptic flow data, an accurate knowledge of the optical potential will,
therefore, be required and the problem of the momentum dependence or independence of the iso-vector potential will have to be resolved.

%\begin{turnpage}
\begin{figure}[tb]
\begin{center}
\begin{minipage}{0.49\textwidth}
\includegraphics[width=18pc]{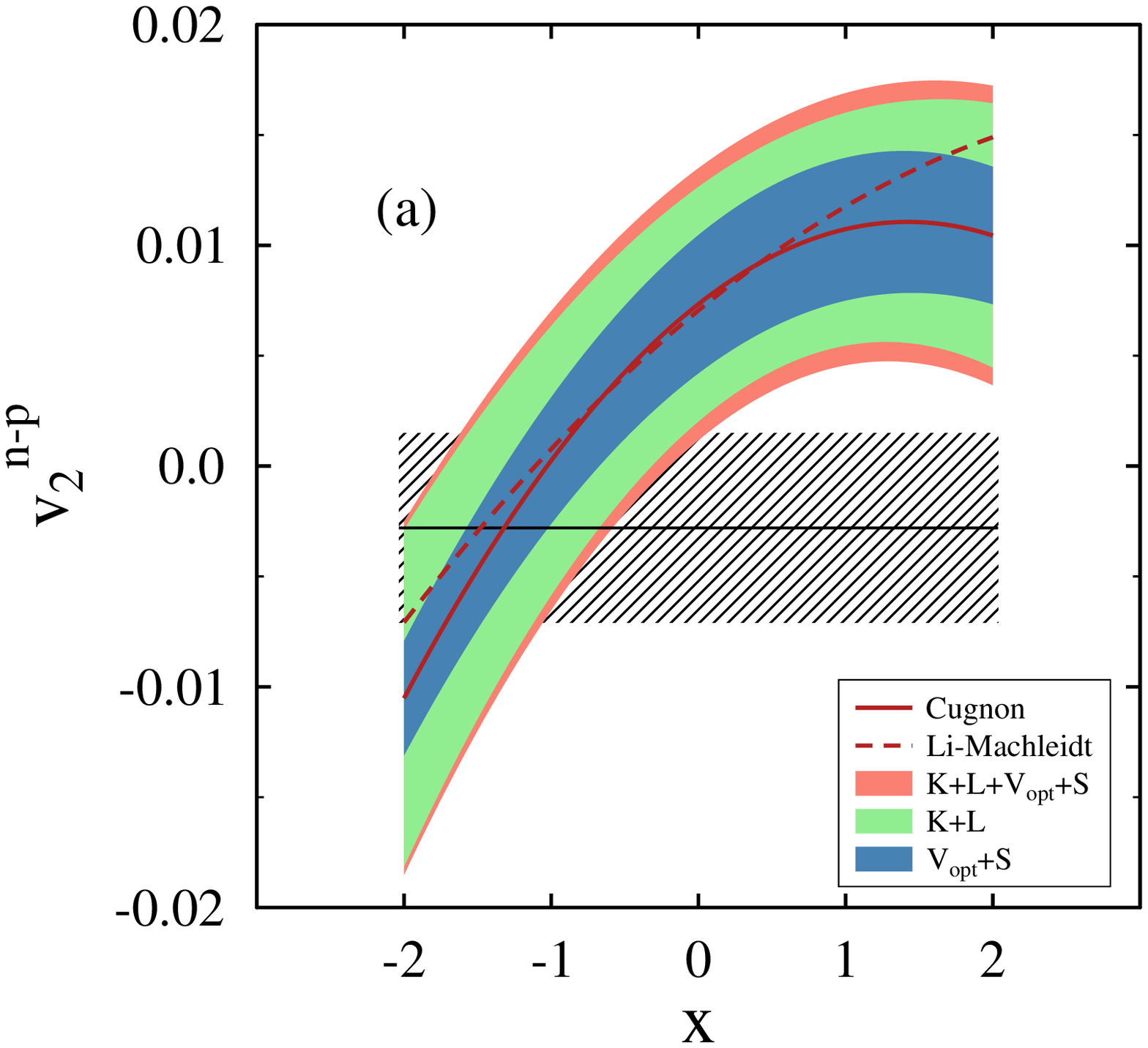}
\end{minipage}
\begin{minipage}{0.49\textwidth}
\includegraphics[width=17.5pc]{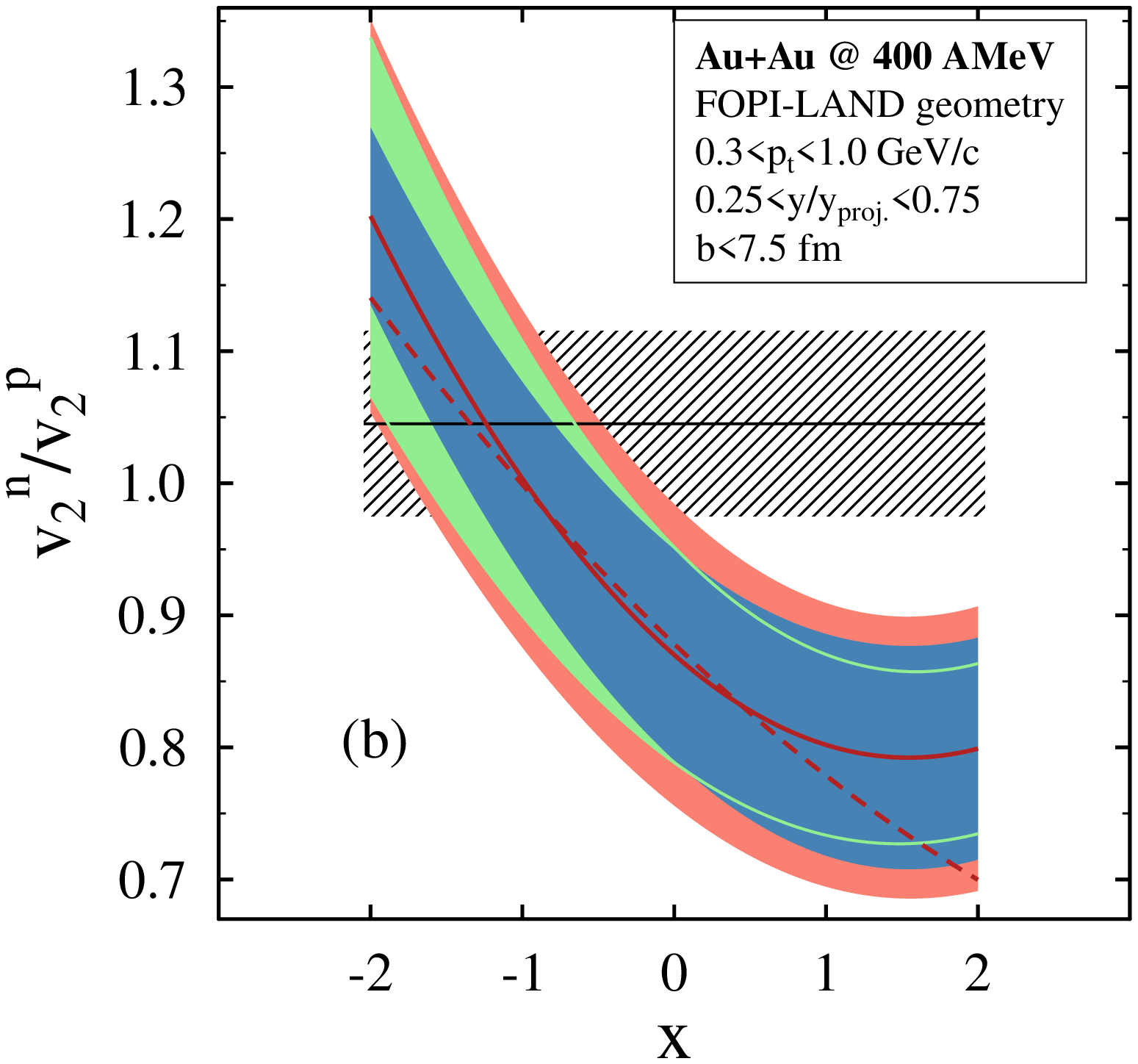}
\end{minipage}
\end{center}
\caption{\figlab{efdefrmoddep}(Color online) Model dependence of npEFD (a) and npEFR (b) and comparison with FOPI-LAND experimental
data, integrated over impact parameter b$\leq$7.5 fm. Sensitivity to the different model parameters, compressibility modulus ($K$), width of nucleon wave function
($L$), optical potential ($V_{opt}$) and parametrization of the symmetry energy ($S$) are displayed. 
The total model dependence is obtained by adding, in quadrature, individual sensitivities.}
\end{figure}
%\end{turnpage}

In~\figref{efdefrmoddep} the model dependence of impact parameter integrated npEFR and npEFD
to variations of the compressibility modulus ($K$), 
width of the nucleon wave function($L$), optical potential ($V_{opt}$) and SE parametrization ($S$) are presented. 
The central curve (full line) was obtained employing the parameter values mentioned above. By using the
vacuum Li-Machleidt nucleon-nucleon cross-sections~\cite{Li:1993ef,Li:1993rwa} instead of the Cugnon ones, 
while keeping the other model parameters unchanged, both npEFD and npEFR remain practically the same irrespective
of the value of $x$. As such the sensitivity to cross-section parametrizations has not been included in the total sensitivity bands.  
The commonly employed value for the compressibility modulus, a soft $K$=210 MeV, has been extracted from the
multiplicity ratio of $K^+$ production in heavy (Au+Au) over light (C+C) nuclei at incident energies
close to 1 AGeV by the KaoS Collaboration~\cite{Sturm:2000dm,Fuchs:2000kp,Hartnack:2005tr} but at 
lower incident energies the situation is not as clear: the KaoS result points to an even softer EoS
while the study of sidewards flow  or stopping by the FOPI collaboration~\cite{Andronic:2003dc,FOPI:2011aa} 
does not exclude a somewhat stiffer isoscalar EoS. For the case of the nucleon wave function width, the
value $L$=4.33 fm$^2$ is commonly employed in transport models simulations of collisions of lighter
nuclei while for the simulation of heavier nuclei an increase to the value $L$=8.66 fm$^2$ is found necessary to prevent
nucleon evaporation. The optical model dependence and SE parametrization
dependence are the same as presented in~\figref{npefdoptpotsens}. 

The variation of the $K$ and $L$ model parameters has been performed within ranges that take into account the facts
mentioned in the previous paragraph together with the requirement that the simulated values for the elliptic flow of
neutrons and protons for $x$=-1 be within 25$\%$ from the experimental ones. This limit was chosen because
 it represents about 3 standard deviations of the combined experimental and numerical uncertainties.
A search for the allowed values for $K$ and $L$ that obey this constraint has been performed with the following outcome. 
Increasing the compressibility modulus
 results in higher elliptic flow absolute values, reaching the upper 25$\%$ off the experimental value boundary in the region
$K$=270$\div$280 MeV. By decreasing the compressibility modulus below $K$=210 MeV a saturation region is reached with values
well within the 25$\%$ off region just below $K$=190 MeV. The dependence of the elliptic flow values on the nucleon wave function
width $L$ proves to be approximately parabolic with a maximum absolute value for the elliptic flows, compatible with the imposed constraint,
 reached close to $L$=3.5 fm$^2$ and crossing the lower 25$\%$ boundary for the values $L$=2.5 fm$^2$ and $L$=7.0 fm$^2$. Consequently,
 to produce the results presented in~\figref{efdefrmoddep} the following variation ranges for the K and L model parameters
have been adopted: $K$=190$\div$280 MeV and $L$=2.5$\div$7.0 fm$^2$.  Increasing the compressibility modulus
to $K$=300 MeV or the nucleon wave function width to $L$=8.66 fm$^2$ produces elliptic flow values that can deviate from the
experimental ones with up to 40-50$\%$ but with marginal impact on the allowed values for the asy-EoS stiffness parameter $x$.

The sensitivities of npEFD and npEFR to model parameter values are similar; both show a model dependence
on the optical potential that is almost independent of the stiffness of asy-EoS while the averaged dependence
on the parametrization of SE is more pronounced for asy-soft scenarios.
The $L$ dependence is slightly more important for an asy-stiff than for an asy-soft EoS for both npEFD and
npEFR. The $K$ dependence is important for both npEFD and npEFR irrespective of the value chosen for the asy-EoS stiffness
 with the exception of the asy-stiff region for npEFD where it represents the most important source of uncertainty. 
The model dependence due to these last two parameters makes up most of the sensitivity of npEFD, 
especially for the case of an asy-stiff scenario. The same holds true for npEFR in the asy-stiff region while
for the case of an asy-soft scenario the importance of the optical potential and SE parametrization is equally or slightly more important.

The total model dependence $K+L+V_{opt}+S$ of both npEFD and npEFR is almost insensitive to the stiffness of asy-EoS and
is comparable in absolute magnitude with the experimental uncertainty of the respective quantity.
 For each observable the experimental value and its uncertainty are depicted
in~\figref{efdefrmoddep} by the horizontal line and hatched band, respectively.
A clear separation of the theoretical and experimental bands,
amounting to about one standard deviation, exists in the super-soft scenario region.
% The separation
%of the theoretical and experimental bands is clear and it amounts to about one standard deviation effect in 
%the super-soft scenario region. 

%Dropping the constraint that the effect of model parameter variations should impact
%elliptic flow values with at most 25$\%$ of their experimental values the separation between theoretical and experimental bands
%drops to about a 2 standard deviations in the same scenario region as before, when the model parameters are varied
%within the presented limits. 

\subsection{Constraints on asy-EoS}

The comparison with the experimental results (\figref{efdefrmoddep}) permits
the extraction of estimates for the stiffness of asy-EoS: x=-1.50$^{+1.75}_{-1.00}$ from npEFD and x=-1.25$^{+1.25}_{-1.00}$ from
npEFR. These constraints translate, using the parabolic expansion of the asy-EoS around the saturation point
\begin{eqnarray}
 E_{sym}(\rho)=E_{sym}(\rho_0)+\frac{L_{sym}}{3}\frac{\rho-\rho_0}{\rho_0}+\frac{K_{sym}}{18}
\frac{(\rho-\rho_0)^2}{\rho_0^2},
\eqlab{lsymksym}
\end{eqnarray}
into the following estimates for the slope and curvature parameters of the symmetry energy: $L_{sym}$=129$^{+46}_{-80}$ MeV, $K_{sym}$=272$^{+291}_{-508}$ MeV (npEFD) and
$L_{sym}$=118$^{+45}_{-57}$ MeV, $K_{sym}$=199$^{+291}_{-362}$ MeV (npEFR). The obtained values for $L_{sym}$ are larger
by a factor of 2 and by 50$\%$ than the ones extracted from an
analysis of neutron skin thickness and isospin diffusion at lower energies~\cite{Tsang:2012se}, respectively. The central values of the npEFD and
npEFR based constraints favor therefore a density dependence of the symmetry energy above the saturation point close to mildly stiff or linear.
They are consistent with each other, the difference between the central values is a consequence of the imperfect
theoretical description of the experimental elliptic flow data at the favored value for the $x$ parameter.
This difference can in principle be eliminated by a finer than here attempted tuning of model parameters.

In~\figref{symenconstraints} the explicit constraints on the density dependence of SE obtained in this study from the
comparison of theoretical and experimental values of npEFD and npEFR are presented. As npEFD and npEFR are not independent
observables, due to the constraint that experimental elliptical flow data be reproduced
at a value of the stiffness parameter $x$ close to the extracted one, only one band, obtained from averaging
the npEFD and npEFR constraints, is advanced for the allowed values for the asy-EoS stiffness. The result
of Ref.~\cite{Russotto:2011hq} is added for comparison. The two studies employ independent flavors of the QMD transport model
(T\"ubingen QMD vs. UrQMD) and parametrizations of isovector EoS that differ: Gogny inspired
(momentum dependent potential) vs. power law (momentum independent potential). 

%Relinquishing the requirement of a close reproduction of
Abandoning the requirement of a close description of
the experimental elliptic flow values for the central estimates allows one to treat npEFD and npEFR as independent observables. 
The extracted constraints for the values of the $x$ parameter are wider if extracted from npEFD and npEFR independently, but the
overlapping region of the two differs only slightly from the one presented in~\figref{symenconstraints}. This brings strong
support to the conclusion that the obtained constraint for the symmetry energy stiffness is model independent by providing evidence
that an asy-EoS stiffness close to one corresponding to the $x$=-1 scenario is favored.% irrespective of wether the experimental elliptic flow values are closely reproduced or not.

The constraints on
the density dependence of SE obtained with these different ingredients are in agreement with each other.
This contrasts with the current status of the effort to constrain the SE from $\pi^-/\pi^{+}$ ratios: a study employing
IBUU transport model and the Gogny inspired asy-EoS~\cite{Xiao:2009zza} favors a soft asy-EoS, a second
study which uses IQMD and the power-law parametrization of SE~\cite{Feng:2009am} points towards a stiff asy-EoS
while the work of~\cite{Xie:2013np} within the Boltzmann-Langevin approach supplemented by a power-law parametrization
of SE concludes that a super-soft scenario is the realistic one. The reason for this strong disagreement has most likely
nothing to do with the parametrization used for the EoS or its momentum (in)dependence but may originate in medium effects
on both $\Delta$ resonance and pion production cross-sections 
(including their isospin dependent energy thresholds) or other related topics~\cite{DiToro:2010ku,Ferrini:2005jw}.

The result presented in this Article is robust, the model dependence of the presented observables, while important, is well
understood and constraints obtained by employing different parametrizations of the asy-EoS are compatible with each other.
An improvement of the current theoretical model is mandatory to allow, together with more accurate experimental data of
elliptic flow of neutrons and protons as expected to be delivered by the ASY-EOS Collaboration~\cite{Russotto:2012jb},
a tighter constraint on the high density dependence of the symmetry energy.
%\begin{turnpage}
\begin{figure}[t]
\includegraphics[width=20pc]{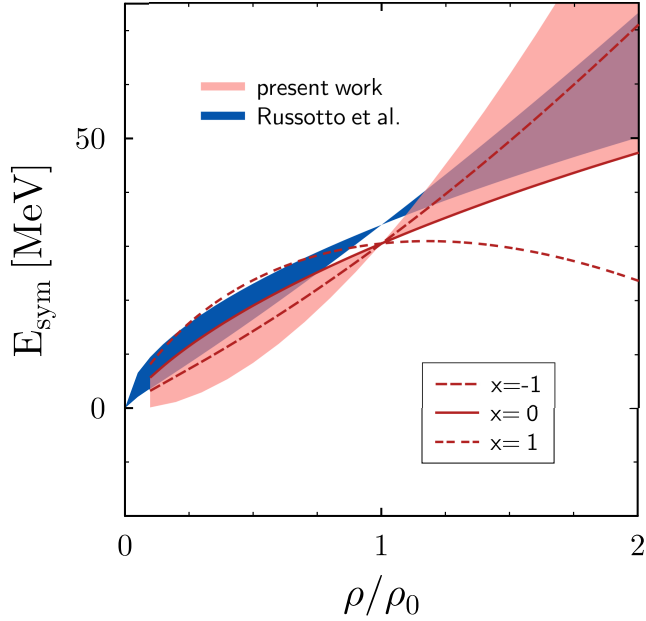}
\caption{\figlab{symenconstraints} (Color online) Constraints on the density dependence of the symmetry energy obtained from
comparing theoretical predictions for npEFD and npEFR to FOPI-LAND experimental data. The result of Russotto~{\it et al.}
~\cite{Russotto:2011hq} is also shown together with the Gogny inspired SE parametrization for three values
of the stiffness parameter: x=-1 (stiff), x=0  and x=1 (soft).}
\end{figure}
%\end{turnpage}

\section{Conclusions}
Constraints on the high density dependence of the symmetry energy (SE) have been extracted by comparing theoretical
predictions of neutron-proton elliptic flow differences (npEFD) and neutron-proton elliptic flow ratios (npEFR)
 with experimental results obtained from a recent analysis of the FOPI-LAND
experimental data for Au+Au collisions at 400 MeV/nucleon. The T\"ubingen QMD model supplemented with
the Hartnack-Aichelin and Gogny parametrizations of the isoscalar EoS and asy-EoS respectively, for the central values, has been employed. 
A thorough study of the model dependence
of npEFD and npEFR has been performed concluding that, while the sensitivity to uncertainties in the model
parameters is important, the two observables offer the opportunity to extract information about the SE
above the saturation point. Furthermore, the results of the present study supplemented with those of Ref.~\cite{Russotto:2011hq}
allow one to conclude that constraints on symmetry energy extracted from elliptic flow data are independent on its parametrization,
suggesting that an almost model independent extraction can be achieved in this case.
This contrasts with the case of $\pi^-/\pi^{+}$ ratios where the stiffnesses of asy-EoS extracted using
different parametrizations for SE or transport models can be extremely different. 

We have imposed that the experimental elliptic flow values of neutrons and protons are reproduced as closely as possible
and accomplished that to within 10-15$\%$ by changing model parameters within limits commonly employed in the literature.
Averaging the constraints extracted independently from npEFD and npEFR one obtains the following allowed values for
the parameters describing the stiffness of the symmetry energy: $L_{sym}$=122$\pm$57 MeV and $K_{sym}$=229$\pm$363 MeV. Together with the estimates obtained
in Ref.~\cite{Russotto:2011hq} we advance the following constraint, obtained from averaging these results,
on the stiffness of asy-EoS:  $L_{sym}$=106$\pm$46 MeV and $K_{sym}$=127$\pm$290 MeV. It corresponds to a 
moderately stiff to linear density dependence and excludes the super-soft and, with a lesser degree of confidence, the soft asy-EoS
scenarios from the list of possibilities. An improvement of the current theoretical model, in the sense of reducing 
theoretical uncertainties, is mandatory to allow, together with more accurate experimental data of elliptic flow of neutrons
and protons, a tighter constraint on the high density-dependence of the symmetry energy.

%\section{Acknowledgments}
\begin{acknowledgments}
The research of M.D.C. has been financially supported by the Romanian Ministry of Education and
Research through contract PN09370103. Q.L. acknowledges financial support from the
National Natural Science Foundation of China (Grant No. 11375062) and
 the Zhejiang Provincial Natural Science Foundation of China (Grant No. Y6090210).
\end{acknowledgments}

\bibliography{references}

%\AtEveryBibitem{% Clean up the bibtex rather than editing it
% \clearfield{eprint}
%}

\end{document}